\documentclass[nofootinbib,showpacs,preprintnumbers,amsmath,amssymb,floatfix]{revtex4}
\usepackage{epsf}

\begin{document}

\title{Singlet structure function $g_1$ at small $x$ and small $Q^2$}

\vspace*{0.3 cm}

\author{B.I.~Ermolaev}
\affiliation{Ioffe Physico-Technical Institute, 194021
 St.Petersburg, Russia}
\author{M.~Greco}
\affiliation{Department of Physics and INFN, University Rome III,
Rome, Italy}
\author{S.I.~Troyan}
\affiliation{St.Petersburg Institute of Nuclear Physics,
188300 Gatchina, Russia}

\begin{abstract}
Explicit expressions for the singlet $g_1$  at small $x$ and small
$Q^2$ are obtained with  the total resummation of the leading
logarithmic contributions. It is shown that $g_1$ practically does
not depend on $x$ in this kinematic region. In contrast, it would
be interesting to investigate its dependence on the invariant
energy $2pq$ because, being $g_1$ positive at small $2pq$, it can
turn negative at greater values of this variable. The position of
the turning point is sensitive to the ratio between the initial
quark and gluon densities, so its experimental detection  would
enable to estimate this ratio.
\end{abstract}

\pacs{12.38.Cy}

\maketitle

\section{Introduction}

The Standard Approach (SA) for the theoretical description of the
spin structure function $g_1$ at large and small $x$ is based on
the Altarelli-Parisi or DGLAP $Q^2-$  evolution equations
\cite{dglap} complemented with global fits \cite{fits} for the
initial parton densities. Originally, SA was suggested for
describing the region of large $x$ but later it has been applied
for investigating the polarized DIS at small $x$ as well. As SA
neglects the total resummation of leading $\ln(1/x)$, which
becomes necessary at small $x$, the singular $(\sim x^{-\alpha})$
factors are introduced in the fits for the initial parton
densities. As it was shown in Refs.~\cite{egt2,egt3}, such factors
act as the leading singularities\footnote{They are simple poles
whereas the total resummation leads to the leading singularity as
the square root branch point.} in the Mellin space. They ensure
the steep rise of $g_1$ at $x \ll 1$ and indeed mimic the impact
of the total resummation of leading $\ln(1/x)$  terms.
Alternatively, when the total resummation  is taken into account,
those singular factors become unnecessary, so the initial parton
densities can be fitted with much simpler expressions. The total
resummation of $\ln(1/x)$ contributions to the anomalous
dimensions and the coefficient functions of the singlet component
of $g_1$ was done in Ref.~\cite{ber} in the Double- Logarithmic
Approximation under the assumption of $\alpha_s$ fixed at an
unknown scale. More precise results including the running
$\alpha_s$ effects were obtained in Ref.~\cite{egt}.

In the present paper, we extend the results of Ref.~\cite{egt} to
consider the small- $x$ behavior of the singlet $g_1$  in more
detail. In particular, we give a special attention to the
kinematic region where not only $x$ but also $Q^2$ are small. On
one hand, this kinematics has been investigated experimentally by
the COMPASS collaboration, see Ref.~\cite{compass}. On the other
hand, the region of small $Q^2$ is clearly beyond the reach of SA.
We show that in this kinematics $g_1$ can be practically
independent of $x$ even for $x \ll 1$. We obtain that $g_1$, being
positive at small values of the invariant energy $2pq$, can turn
negative when $2pq$ increases. The position of the turning point
is sensitive to the ratio between the initial quark and gluon
densities. Then we also show that, in spite of the presence of
large factors providing $g_1$ with the Regge behavior at small
$x$, the interplay between initial quark and gluon densities might
keep $g_1$ at small $x$ close to zero even at small $x$,
regardless of values of $Q^2$.

The paper is organized as follows: in Sect.~2 we remind and
explain the basic formulae for $g_1$ singlet obtained in
Ref.~\cite{egt}. These formulae include the total resummaion of
the leading logarithms of $x$. In our approach, the coefficient
functions for $g_1$ are expressed through new anomalous
dimensions.  Explicit expressions for them are presented in
Sect.~3. We focus on $g_1$ at small $Q^2$ in Sect.~4. As our
approach is perturbative, we are interested in minimizing the
influence of non-perturbative contributions. To his aim we
introduce an optimal mass scale in Sec.~5. The asymptotics of
$g_1$ at small $x$ is considered in Sect.~6. Suggestions for new
simple fits for the initial parton densities at small $Q^2$ are
briefly discussed in Sec.~7.  Sect.~8 contains our numerical
results, and finally Sect.~9 is for the concluding remarks.

\section{Expressions for $g_1$ at large $Q^2$}

The singlet structure function $g_1$ at small $x$ was studied in
Ref.~\cite{egt}.
According to it, $g_1$ can be represented in the form of the Mellin integral:

\begin{equation}
\label{g1int}
g_1(x,Q^2) =  \frac{<e^2_q>}{2}
\int_{- \imath \infty}^{\imath \infty} \frac{d \omega}{2 \pi \imath}
\Big(\frac{1}{x} \Big)^{\omega}
\Big[ \Big(C_q^{(+)}e^{\Omega_{(+)} y} +
C_q^{(-)}e^{\Omega_{(-)} y} \Big) \delta q +
\Big(C_g^{(+)}e^{\Omega_{(+)} y} +
C_g^{(-)}e^{\Omega_{(-)} y} \Big) \delta g
\Big]
\end{equation}
where $<e^2_q>$ stands for the sum of electric charges: $<e^2_q> =
10/9$ for $n_f = 4$, $y = \ln(Q^2/\mu^2)$, with $\mu$ being the
starting point of the $Q^2$- evolution, $\delta q$ is the initial
averaged quark density: $<e^2_q> \delta q = e^2_u \delta u + e^2_d
\delta d +...$ whereas $\delta g$ is the initial gluon density.

The other ingredients of the integrand in Eq.~(\ref{g1int}) are
expressed in terms of the anomalous dimensions $H_{ik}$, with $i,k
= q,g$. The exponents $\Omega_{(\pm)}$ and coefficient functions
$C^{(\pm)}_{q,g}$ are:
\begin{equation}
\label{omegapm} \Omega_{(\pm)} = \frac{1}{2} \big[ H_{qq} + H_{gg}
\pm R \big]~,
\end{equation}
\begin{eqnarray}\label{cbk}
C_q^{(+)} &=& \frac{\omega}{RT} \Big[
(H_{qq}-\Omega_{(-)})(\omega-H_{gg}) + H_{qg}H_{gq} +
H_{gq}(\omega-\Omega_{(-)}) \Big]~, \\ \nonumber C_q^{(-)} &=&
\frac{\omega}{RT} \Big[ (\Omega_{(+)}-H_{qq})(\omega-H_{gg}) -
H_{qg}H_{gq} + H_{gq}(\Omega_{(+)}-\omega) \Big]~, \\ \nonumber
C_g^{(+)} &=& \frac{\omega}{RT} \Big[
(H_{gg}-\Omega_{(-)})(\omega-H_{qq}) + H_{qg}H_{gq} +
H_{qg}(\omega-\Omega_{(-)}) \Big]
\Big(-\frac{A'}{2\pi\omega^2}\Big)~, \\ \nonumber C_g^{(-)} &=&
\frac{\omega}{RT} \Big[ (\Omega_{(+)}-H_{gg})(\omega-H_{qq}) -
H_{qg}H_{gq} + H_{qg}(\Omega_{(+)}-\omega) \Big]
\Big(-\frac{A'}{2\pi\omega^2}\Big)~.
\end{eqnarray}
Here
\begin{equation}\label{rt}
R= \sqrt{ (H_{qq} -  H_{gg})^2 + 4 H_{qg} H_{gq}}~,\qquad T=
\omega^2 - \omega (H_{gg} + H_{qq}) + (H_{gg}H_{qq} -
H_{gq}H_{qg})~
\end{equation}
and
\begin{equation}\label{aprime}
A'(\omega) = \frac{1}{b} \Big[ \frac{1}{\eta} - \int_{0}^{\infty}
 \frac{d \rho e^{- \omega \rho}}{(\rho + \eta)^2} \Big]~
\end{equation}
with $\eta=\ln(\mu^2/\Lambda_{QCD}^2)$ and $b=(33-2 n_f)/(12\pi)$.
The additional factor $\Big(-\frac{A'}{2\pi\omega^2}\Big)$ in the
coefficients $C_g^{(\pm)}$ is the small-$\omega$ estimate for the
quark box which relates the initial gluons to the electromagnetic
current. $A'(\omega)$ stands for the QCD coupling $\alpha_s$ in
the box in the Mellin space.

\section{anomalous dimensions}

The anomalous dimensions $H_{ik}$ obey the following system of
equations:

\begin{eqnarray}
\label{eqhik}
\omega H_{qq} &=&  b_{qq} +  H_{qg} H_{gq} +  H^2_{qq} \\ \nonumber
\omega H_{gg} &=&  b_{gg} +  H_{gg} H_{qg} +  H^2_{gg} \\ \nonumber
\omega H_{qg} &=&  b_{qg} +  H_{qg} H_{gg} +  H_{qg}H_{gg} \\ \nonumber
\omega H_{gq} &=&  b_{gq} +  H_{gq} H_{qq} +  H_{gg}H_{gq}
\end{eqnarray}
where
\begin{equation}
\label{bik}
b_{ik} = a_{ik} + V_{ik}~,
\end{equation}
with the Born contributions $a_{ik}$ defined as follows:

\begin{equation}\label{aik}
a_{qq} = \frac{A(\omega) C_F}{2 \pi}~,\quad a_{qg} = \frac{
A'(\omega) C_F}{\pi}~,\quad a_{gq} = - \frac{n_f A'(\omega)}{2
\pi}~,\quad a_{gg} = \frac{4 N A(\omega)}{2 \pi}~.
\end{equation}
$A'$ is given by the Eq.(\ref{aprime}) and
\begin{equation}\label{a}
A(\omega) = \frac{1}{b} \Big[ \frac{\eta}{\eta^2 + \pi^2} -
\int_{0}^{\infty} \frac{d \rho e^{- \omega \rho}}{(\rho + \eta)^2
+ \pi^2} \Big]~
\end{equation}
is the Mellin representation of the QCD running coupling
$\alpha_s$ involved in the quark-gluon ladder, with the proper
account of its analytic properties. In Eq.(\ref{aik}) we use the
standard notations for $C_F=(N^2-1)/(2N)=4/3$ and $N=3$.

Finally,
\begin{equation}
\label{vik} V_{ik} = \frac{m_{ik}}{\pi^2} D(\omega)~,
\end{equation}
where
\begin{equation}
\label{mik}
m_{qq} = \frac{C_F}{2 N}~,\quad
m_{gg} = - 2N^2~,\quad
m_{gq} = n_f \frac{N}{2}~,\quad
m_{qg} = - N C_F~,
\end{equation}
and
\begin{equation}
\label{d}
D(\omega) = \frac{1}{2 b^2}
\int_{0}^{\infty} d \rho e^{- \omega \rho}
\ln \big( (\rho + \eta)/\eta \big)
\Big[ \frac{\rho + \eta}{(\rho + \eta)^2 + \pi^2} + \frac{1}{\rho + \eta}\Big]~
\end{equation}
is the factor that accounts for non-ladder diagrams.

The solution to Eq.~(\ref{eqhik}) is
\begin{eqnarray}\label{hik}
&& H_{qq} = \frac{1}{2} \Big[ \omega + Z + \frac{b_{qq} -
b_{gg}}{Z}\Big],\qquad H_{qg} = \frac{b_{qg}}{Z}~, \\ \nonumber &&
H_{gg} = \frac{1}{2} \Big[ \omega + Z - \frac{b_{qq} -
b_{gg}}{Z}\Big],\qquad H_{gq} =\frac{b_{gq}}{Z}~
\end{eqnarray}
where
\begin{equation}
\label{z}
 Z = \frac{1}{\sqrt{2}}\sqrt{(\omega^2 - 2(b_{qq} + b_{gg})) +
\sqrt{(\omega^2 - 2(b_{qq} + b_{gg}))^2 - 4 (b_{qq} - b_{gg})^2 -
16b_{gq} b_{qg} }}~.
\end{equation}

\section{Expressions for the singlet $g_1$ at small $Q^2$}

Eq.~(\ref{g1int}) states that $g_1$ does not depend on $Q^2$ when
$Q^2 \sim \mu^2$. In this case $g_1$ depends only on $z \equiv
\mu^2/(2pq)$:

\begin{equation}
\label{g1q0} g_1(z) =  \frac{<e^2_q>}{2} \int_{- \imath
\infty}^{\imath \infty} \frac{d \omega} {2 \pi \imath}
\Big(\frac{1}{z}\Big)^{\omega} \Big[ \omega \frac{ \omega - H_{gg}
+ H_{gq}}{T} \delta q + \omega \frac{ \omega - H_{qq} + H_{qg}}{T}
\Big(-\frac{A'}{2\pi\omega^2}\Big) \delta g \Big]~.
\end{equation}
Eq.~(\ref{g1int}) was obtained for  $Q^2 \gtrsim \mu^2$ and cannot
be used for studying the $Q^2$ -dependence of $g_1$ at $Q^2 <
\mu^2$. However,
it would be interesting to extend our approach to this region. In
order to do so, we suggest to modify Eq.~(\ref{g1int}), replacing
$Q^2$  by $(Q^2 + \mu^2)$. Although such a shift takes us out of
the logarithmic accuracy we have always kept in our approach, it
looks quite reasonable and natural. Indeed, let us in the first
place consider a contribution of a ladder Feynman graph at $x \ll
1$ with the DL accuracy. The graph includes the quark and gluon
rungs. Integrations of the quark rungs are infrared-stable, being
regulated with the quark mass $m_q$. On the contrary, integrations
of the gluon rungs are IR-divergent, so they must be regulated.
The standard way of the IR-regulating in QED and QCD is providing
gluons with a mass $\mu$ which acts as an IR cut-off. It is also
convenient to choose $\mu \gg m_q$ and replace $m_q$ by $\mu$ in
the quark propagators as was first suggested in Ref.~\cite{kl}.
After that $m_q$ can be dropped. Now both gluon and quark rungs of
the ladder are IR-stable and $\mu$ -dependent.   The
simplification of the spin structure can be done with the standard
means (see e.g. the review \cite{g}). It is appropriate to use the
standard Sudakov variables for integrations over momenta $k_i$ of
ladder virtual quarks and gluons: $k_i = \alpha_i (p -
(m_q^2/2pq)) + \beta_i (q + x p)+ k_{\perp}$. After that the DL
contribution of a ladder graph with $n$ rungs is proportional to
the integral $J_n$:

\begin{eqnarray}
\label{ladder} J_n =  \int \frac{d k^2_{n~\perp} d \alpha_n d
\beta_n \delta (w \beta_n - Q^2 - \mu^2 + w\alpha_n \beta_n -
k^2_{n~\perp})}{ \alpha_n \beta_n - k^2_{n~\perp}/w  -\mu^2/w} \\
\nonumber \int \frac{d k^2_{n-1~\perp} d \alpha_{n-1} d
\beta_{n-1} \delta (w\alpha_n \beta_{n-1} - \mu^2 - (k^2_{n~\perp}
+k^2_{n-1~\perp}
))}{w \alpha_{n-1} \beta_{n-1} - k^2_{n-1~\perp}/w -\mu^2/w}~... \\
\nonumber \int \frac{d k^2_{1~\perp} d \alpha_1 d \beta_1 \delta
(-w \alpha_1 - \mu^2 + w\alpha_1 \beta_1 - k^2_{1~\perp})}{
\alpha_1 \beta_1 - k^2_{1~\perp}/w -\mu^2/w}
\end{eqnarray}
where the rungs are numbered from the bottom  to the top of the
ladder. We have used the notation $w \equiv 2pq$. As we consider
$x \ll 1$, we neglected the term $-w x \alpha_n$ coming from the
representation $2qk_n = w\beta_n -w x \alpha_n$ in the argument of
the first $\delta$ -function and similar terms in the other
$\delta$ -functions. Eq.~(\ref{ladder}) manifests that the $Q^2$
-dependence in $J_n$ at $x \ll 1$ is given by the term  $Q^2 +
\mu^2$ in the first $\delta$ -function only. Neither accounting
for non-ladder graphs nor
 accounting for single logarithms, including the running $\alpha_s$ effects,
 change this situation. So, the replacement
\begin{equation}\label{shift}
Q^2 \to \widetilde{Q}^2 \equiv Q^2  + \mu^2~,
\end{equation}
is confirmed by the analysis of the structure of the Feynman
graphs, though as  the replacement is beyond the logarithmic
accuracy, it can be called model-dependent.

Further, DGLAP exploits the $Q^2$ -evolution, with $Q^2$ being the
upper limit of integrations over $k^2_{r~\perp}~(r = 1,2,..)$, so
the first loop DGLAP double-logarithmic contribution $J_1^{DGLAP}
= \ln(Q^2/\mu^2)$. This contribution is large only when $Q^2 \gg
\mu^2$.  In contrast, we use the evolution with respect to
$\mu^2$, the integrations over $k^2_{r~\perp}$ run up to $w$
instead of $Q^2$, so our approach is not restricted by the region
of large $Q^2$. For example, when $n = 1$, Eq.~(\ref{ladder})
yields
\begin{equation}\label{rung1}
J_1 = \ln (w/\mu^2)~,
\end{equation}
so $J_1$ does not depend on $Q^2$ at all. The $Q^2$ -dependence
appears in the next loops. In particular, when $n=2$,
\begin{equation}\label{rung2}
J_2 = -(1/2)\ln^2 (w/\mu^2) \ln(\widetilde{Q}^2/w) +
(1/6)\ln^3(\widetilde{Q}^2/w)~,
\end{equation}
so it depends on $Q^2$ through $Q^2 + \mu^2$. It agrees with
Eq.~(\ref{shift}). In contrast to DGLAP, double logarithms in
Eqs.~(\ref{rung1},\ref{rung2}) do not disappear when $Q^2 \to 0$.

Using Eq.~(\ref{shift}) makes possible to rewrite
Eq.~(\ref{g1int}) in the form
  convenient equally for large and small $Q^2$:
\begin{eqnarray}
\label{g1smallq} g_1(x, Q^2) = \frac{<e^2_q>}{2} \int_{- \imath
\infty}^{\imath \infty} \frac{d \omega}{2 \pi \imath}
 \Big(\frac{1}{z + x}
\Big)^{\omega} \Big[ \Big(C_q^{(+)}(\omega) \Big(\frac{Q^2 +
\mu^2}{\mu^2} \Big)^{\Omega_{(+)}} + C_q^{(-)}(\omega)
\Big(\frac{Q^2 + \mu^2}{\mu^2} \Big)^{\Omega_{(-)}}\Big) \delta q
\\ \nonumber  -\frac{A'}{2 \pi \omega^2} \Big(C_g^{(+)}(\omega)\Big(\frac{Q^2 + \mu^2}{\mu^2}
\Big)^{\Omega_{(+)}} + C_g^{(-)}(\omega)\Big(\frac{Q^2 +
\mu^2}{\mu^2} \Big)^{\Omega_{(-)}}\Big)\delta g \Big]~.
\end{eqnarray}
Eq.~(\ref{g1smallq}) coincides with Eq.~(\ref{g1int}) when $Q^2
\gg \mu^2$ and also reproduces  Eq.~(\ref{g1q0}) when $Q^2 = 0$.
 Eq.~(\ref{g1smallq}) shows that the
$x$- and $Q^2$- dependence of $g_1$ are getting weaker with
decreasing $Q^2$ so that $g_1$ at $Q^2 \ll \mu^2$ depends rather
on $z = \mu^2/(2pq)$ than on $x$ or $Q^2$.
Eqs.~(\ref{g1int},\ref{g1smallq}) describe the leading $Q^2$
-dependence of $g_1$ at $Q^2 \gg \mu^2$. Similarly,
Eq.~(\ref{g1smallq}) describes the leading $Q^2$ -dependence at
$Q^2 \ll \mu^2$: although logarithms  of $\big((Q^2 +
\mu^2)/\mu^2\big)$ are small here, they are multiplied by leading,
double logarithms of $1/z$ contrary to other, unaccounted $Q^2$
-terms. In what follows we will not discuss the $Q^2$- dependence
of $g_1$ at small $Q^2$ in detail. Instead, we focus on
investigating $g_1$ at $Q^2 \to 0$ where $g_1$ is given by any of
Eqs.~(\ref{g1q0},\ref{g1smallq}). According to
Eqs.~(\ref{g1int},\ref{g1q0},\ref{g1smallq}), the total
resummation of double-logarithms for $g_1$ makes it depend on the
value and the way of introducing the cut-off $\mu$. This
dependence will vanish when the probabilities to find a polarized
quark and gluon are calculated and used in expressions for $g_1$
 instead of the phenomenological
initial densities $\delta q$ and $\delta g$.

\section{Optimal scale for $\mu$}

In order to estimate $\mu$, we discuss below the restrictions for
it. From
\begin{equation}
\label{alpha} \alpha_s(k^2) = \frac{1}{b \ln(k^2/
\Lambda_{QCD}^2)}~,
\end{equation}
 as $k^2 \gg \Lambda_{QCD}^2$ and $k^2 > \mu^2$,
we obtain the obvious restriction for the value of $\mu$:
\begin{equation}
\label{mulambda}
\mu^2 \gg \Lambda_{QCD}^2~.
\end{equation}
Then, the DL contributions from ladder quark rungs  are infrared-
stable, with logarithms there containing masses $m_q$ of the
involved quarks in denominators. In order to calculate ladder
fermion graphs with Infra-Red Evolution Equations  these logs
should be regulated with the infrared cut-off $\mu$. It brings the
second restriction for $\mu$:

\begin{equation}
\label{mumq}
\mu > m_q~.
\end{equation}

Basically, there are no other restrictions for $\mu$. However,
some additional information of $\mu$ comes from the small-$x$
asymptotics of $g_1$. In  Ref.~\cite{egt} it was shown that

\begin{equation}
\label{as}
g_1 \sim (1/x)^{\omega_0}
\end{equation}
when $x \to 0$. It turned out that $\omega_0$ depends on $\eta =
\ln(\mu^2/ \Lambda^2_{QCD})$, in such a way that  $\omega_0$  is
maximal at $\eta = \eta_S \approx 7.5$ with

\begin{equation}
\label{deltas} \omega_0(\eta_S) \equiv \Delta_S \approx 0.86.
\end{equation} We have called $\Delta_S$ the intercept of the singlet
$g_1$. This value is in agreement  with the analysis of
experimental data\cite{koch}. Assuming $\Lambda_{QCD} = 0.1$~GeV
leads to the estimate
\begin{equation}
\label{mu0} \mu_S = \Lambda_{QCD} e^{3.75} \approx 5.5~GeV~.
\end{equation}
On the other hand, from physical considerations, the intercept
$\Delta_S$  should be a constant and should not depend on $\mu$.
This dependence is  the artefact of our approach: we account for
perturbative contributions to the asymptotics and leave out a
possible impact of non-perturbative ones. Taken together, the
perturbative and non-perturbative contributions would make
$\omega_0$ to be $\mu$ -independent. We suggested in
Ref.~\cite{egt} that non-perturbative contributions to  $\omega_0$
happened to be minimal at $\eta = \eta_S$ so in order to minimize
the impact of (the basically unknown) non-perturbative
contributions on $g_1$, we should fix $\mu = \mu_S = 5.5~GeV$. We
call $\mu_S$ the Optimal Scale for the singlet $g_1$. We expect
that  choosing this scale for
 $\mu$ would bring a better agreement
between experimental data and our formulae than other values of
$\mu$. We also suggest that the initial parton densities can be
fitted mostly simply when $\mu$ is fixed at the Optimal Scale. It
is worth of mentioning that in Refs.~\cite{egtns} the Optimal
Scale $\mu_{NS}$ for the non-singlet component of $g_1$ is 5 times
smaller: $\mu_{NS} = 1$~GeV.

\section{Small- $x$ asymptotics of $g_1$}

Before performing numerical analysis of Eq.~(\ref{g1smallq}), it
could be instructive to consider its small-$x$ asymptotics. This
asymptotics is different for small and large $Q^2$ and when $x+z
\to 0$, we obtain
\begin{equation}\label{g1asq}
g_1 \sim \Big( \frac{1}{x+z}\Big)^{\Delta_S}
\frac{K}{\ln^{3/2}1/(x+z)} \Big( \frac{Q^2+\mu^2}{\mu^2}
\Big)^{\Delta_S/2}\Big(\frac{2}{\Delta_S}+\ln\frac{Q^2+\mu^2}{\mu^2}\Big)~
\Big[C_q^{as}\delta q + C_g^{as}\delta g\Big]~,
\end{equation}
where the intercept $\Delta_S$ is given by Eq.~(\ref{deltas}) are
taken at the intercept point $\omega=\Delta_S$.,  $\Delta_S
\approx 0.86$,
\begin{equation}\label{g1asqK}
K = \sqrt{\frac{\widetilde{\Delta_S}}{8\pi}}~,\quad
\widetilde{\Delta_S} = \Delta_S -
\partial [(b_{gg}+b_{qq}) - r]/ \partial \omega~,\quad r
=\sqrt{(b_{gg}-b_{qq})^2+4b_{qg}b_{gq}}~,
\end{equation}
and
\begin{equation}\label{g1asqC}
C_q^{as}=1+\frac{b_{qq}-b_{gg}+2b_{gq}}{r}~,\qquad
C_g^{as}=\Big(1+\frac{b_{gg}-b_{qq}+2b_{qg}}{r}
\Big)\Big(-\frac{A'(\Delta_S)}{2\pi\Delta_S^2}\Big)~,
\end{equation}
with all $b_{ij}$ and their $\omega$-derivatives  in
Eq.~(\ref{g1asqK}) are taken at the intercept point
$\omega=\Delta_S$. The initial parton densities $\delta q
(\omega)$ and $\delta g (\omega)$ are also fixed at $\omega =
\Delta_S$.

When $z \to 0$ and $Q^2 \ll \mu^2$, Eq.~(\ref{g1asq}) turns (we
drop here the unessential overall factor) to
\begin{equation}
\label{g1assmallq} g_1 \sim S(\Delta_S) \delta q (\Delta_S)
\Big(\frac{1}{z}\Big)^{\Delta_S}/\ln^{3/2}(1/z)
\end{equation}
with
\begin{equation}
\label{predexp} S(\Delta_S) = -1 - 0.064~ \delta
g(\Delta_S)/\delta q(\Delta_S)
\end{equation}
Eqs.~(\ref{g1assmallq}-\ref{predexp}) show that the aymptotics of
$g_1$ does not depend on $x$ and $Q^2$ in the small-$Q^2$ region
and the sign of $g_1$ is determined by $S(\Delta_S)$. There can be
three options :

\subsection{Large  and negative $\delta g $:~ $S(\Delta_S) > 0$.}
When the initial gluon density is negative and large so that
\begin{equation}
\label{g1pos} \delta g  <  - 15.64 \delta q~,
\end{equation}
the asymptotics of $g_1$ is positive. It is known that  $g_1 > 0$
at large $z$ where it is given by its Born expression. Therefore,
if $\delta q$ and $\delta g$ are related by Eq.~(\ref{g1pos}),
$g_1$ is positive in the whole range of $z$.

\subsection{Positive or small and negative $\delta g $:~ $S(\Delta_S) < 0$.}
On the contrary, when
\begin{equation}
\label{g1neg} \delta g  > - 15.64 \delta_q~,
\end{equation}
$g_1$, being positive at large $x$, should pass through the zero
value and changes its sign at asymptotically small $z$.

\subsection{Fine tuning:~ $S(\Delta_S) = 0$}
Finally, there might be a strong correlation between $\delta q$
and $\delta g$ :
\begin{equation}
\label{g1zero} \delta g =  15.64 \delta q
\end{equation}
when $z \to 0$. In this case  $g_1$ is positive at large $x$ and
then $g_1 \to 0$ in spite of the large power-like factor
$(1/z)^{\Delta_S}$ in Eq.~(\ref{g1assmallq}).

\section{Fits for initial parton densities}

In the Standard Approach, the initial parton densities $\delta
q(x),~ \delta g(x) $ are fitted from experimental data at $x \sim
1$ and $Q^2 = \mu^2 \approx 1$~GeV$^2$. Then they are evolved with
the anomalous dimensions into the region of large $Q^2:~Q^2 \gg
\mu^2$ and finally evolved with the coefficient functions into the
region of $x \ll 1$. As the coefficient functions in the SA do not
include the total resummation of $\ln x$ and therefore cannot
provide $g_1$ with the steep rise at small $x$, this role is
assigned, in the standard fits\cite{fits}, to the singular factors
$x^{-\alpha}$ which mimic the resummation. In other words, the
impact of the NLO terms in the DGLAP coefficient functions on the
small-$x$ behavior of $g_1$ is actually negligibly small compared
to the impact of the fit. When the resummation is accounted for,
the singular factors can be dropped out and the fits can be
simplified down to expressions $\sim N_{q,g}(1 + c_{q,g} x)$.
Obviously, the straightforward evolution of the fits backwards, to
the region of $Q^2 \ll \mu^2$ is beyond SA. We suggest that the
analyses of the large $Q^2$- and small $Q^2$- experimental data
would be more consistent when the argument $x$ in the new fits is
replaced by $(z + x)$. This argument $\approx x$ at large $Q^2$
and $\approx z$ at small $Q^2$. It means that at small $Q^2$ the
fits should depend on $2pq$ only. We suggest that the fits for
$\delta q(z + x),~ \delta g(z + x)$ can be chosen as linear forms
$\sim N_{q,g}(1 + a_{q,g} (z + x))$, with parameters
$N_{q,g},~a_{q,g}$ fitted from experimental data either at small
or large $Q^2$. However, in the present paper we do not plan to
study new fits in detail. Instead, we assume that after the total
resummations of $\ln x$ has been accounted for, one can
approximate the initial parton densities by constants:
\begin{equation}
\label{fits} \delta q = N_q~,\qquad\delta g = N_g~.
\end{equation}
The analysis of the data on the non-singlet $g_1$ shows that $N_q$
should be positive whereas the sign of $N_g$ is basically unknown.

\section{Numerical results for $g_1$ at  $Q^2 = 0$}

\begin{figure}
\begin{center}
\begin{picture}(240,140)
\put(0,0){
\epsfbox{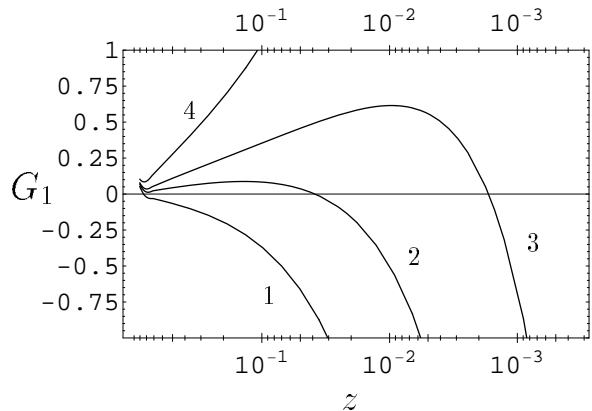}}
\end{picture}
\end{center}
\caption{$G_1$ evolution with decreasing $z=\mu^2/2(pq)$ for
different values of ratio $r= \delta g/\delta q$: curve 1 - for
$r=0$, curve 2 - for $r= -5$~, curve~3 -for $r = -8$ and curve~4
-for $r = -15$.} \label{fig1}
\end{figure}

Both Eq.~(\ref{g1smallq}) and the asymptotic expression
Eq.~(\ref{g1assmallq}) imply that the $x$- dependence of $g_1$ at
small $Q^2$ should be weak even for very small $x$. As a matter of
fact, $g_1$ in this region depends on $2pq$ only. so a plot of
$g_1$ vs $x$ should reveal a very flat behavior, with the
magnitude, $g_1(z)$, depending on the interplay between $\delta q$
and $\delta g$ at different $z = \mu^2/(2pq)$. Presuming that
$\delta q >0$ and approximating $\delta q$ and $\delta g$ by the
constants $N_{q,g}$, we rewrite Eq.~(\ref{g1smallq}) at $Q^2 = 0$
as
\begin{equation}\label{g1num}
g_1(z) = (<e^2_q>/2) N_q G_1(z)
\end{equation}
and calculate $G_1$ numerically. The results for different values
of the ratio $r=N_g/N_q$, $G_1$ are plotted in
Fig.~\ref{fig1}\footnote{ We  remind  that Eq.~(\ref{g1smallq}) is
becoming unreliable when $z \sim 1$  and  should be modified
according to the  prescription of Ref.~\cite{egt}  in order to
describe this region.}. When the  gluon density is neglected, i.e.
$N_g = 0$ (curve~1), $G_1$ being positive at $x \sim 1$, is
getting negative very soon, at $z < 0.5$ and falls fast with
decreasing $z$. When $N_g/N_q = -5$ (curve~2), $G_1$ remains
positive and not large until $z \sim 10^{-1}$, turns negative at
$z \sim 0.03$ and falls afterwards rapidly with decreasing $z$ .
This turning point where $G_1$  changes  its sign is very
sensitive to the magnitude of the ratio $r$~. For instance, at
$N_g/N_q = -8$ (curve~3), $G_1$ passes through zero at $z \sim
10^{-3}$. When $N_g/N_q < -10$, $G_1$ is positive at any
experimentally reachable $z$ (curve~4)~. Therefore, the
 experimental measurement of the turning point
would allow to draw conclusions on the interplay between the
initial quark and gluon densities.

\section{Conclusion}

We have shown that the  study of $g_1$ at small-$Q^2$ could be
 as interesting as in the large-$Q^2$ region. Eq.~(\ref{g1smallq})
 describes the singlet $g_1$ at small $x$ and arbitrary values of
$Q^2$, generalizing both Standard Approach and our previous
results. It accounts for the total resummation of the leading
logarithms of $x$ and $z$ ($z = \mu^2/2pq$). It makes possible to
simplify the fits for the initial parton densities. In general,
$g_1$ includes both perturbative and non-perturbative
contributions. In order to minimize the impact of the latter, we
have introduced an Optimal Scale for $\mu$. In the small- $Q^2$
region, Eq.~(\ref{g1smallq}) predicts that $g_1$  essentially
depends on $2pq$ only and practically does not depend on $Q^2$ and
$x$ even at $x \ll 1$, making the investigation of the $x$
-dependence uninteresting. On the contrary, the study of  the $z$
-dependence of $g_1$ at small $Q^2$ would be useful. Indeed, the
sign of $g_1$ is positive at $z$ close to 1 and can remain
positive or become negative at smaller $z$, depending on the ratio
between $\delta g$ and $\delta q$. Our numerical results are
plotted in Fig.~1.  The position of this point is sensitive to the
ratio $\delta g/\delta q$, so the experimental measurement of this
point would enable to estimate the impact of $\delta g$.

\section{Acknowledgements}
We are grateful to R.~Windmolders who drew our attention to the
problem of describing the $g_1$ singlet in the kinematic region of
small $x$ and $Q^2$. We are also grateful to G.~Altarelli for
useful discussions. The work is partially supported by the Russian
State Grant for Scientific School RSGSS-5788.2006.2

\end{document}